\documentclass[aps,prl,twocolumn,groupedaddress,longbibliography]{revtex4-1}
\usepackage{multimedia,array}
\usepackage{ulem}
\usepackage{xcolor}
\usepackage{appendix}

\newcommand{\ve}{j_{\rm edge}}

\definecolor{colortodo}{RGB}{255,0,0}

\usepackage{graphicx,setspace}
\usepackage{amssymb,amsfonts,amsmath,times,color}
\usepackage[english]{babel}

\begin{document}
\title{Density Mediated Spin Correlations Drive Edge to Bulk Flow Transition in Active Chiral Matter}
\author{Alexander P. Petroff, Christopher Whittington, Arshad Kudrolli} 
\affiliation{Department of Physics, Clark University, Worcester, MA 01610, USA} 

\date{\today}

\begin{abstract}
We demonstrate that edge currents develop in active chiral matter---composed of spinning 
disk-shaped grains with chirally arranged tilted legs confined in a circular vibrating chamber---
due to boundary shielding over a wide range of densities 
corresponding to a gas, fluid, and crystal. The edge currents are then shown to increasingly drive circulating bulk flows with area fraction $\phi$ due to increasing spin-coupling between neighbors mediated by frictional contacts, as percolating clusters develop. Edge currents are observed even in the dilute limit. While, at low $\phi$, the average flux vanishes except within a distance of a single particle diameter of the boundary, the penetration depth grows with increasing $\phi$ till a solid body rotation is achieved corresponding to the highest packing, where the particles are fully caged with hexagonal order and spin in phase with the entire packing. A coarse-grained model, based on the increased collisional interlocking of the particles with $\phi$ and the emergence of order, captures the observed flow fields.
\end{abstract}
\maketitle

Chiral active matter is composed of particles or organisms that intrinsically 
spin~\cite{Tsai2005,Petroff2015,van_Zuiden2016,Dasbiswas2018,liebchen2022chiral,Tan2022}. These materials are naturally out of equilibrium; energy and rotation are
constantly supplied to the system on the scale of the particle, and dissipated by the global motion of
the particles.
The intrinsic rotation of each particle causes colliding particles to
rotate about one another in a preferred direction. Consequently,
locally increasing particle density also increases the local
vorticity. The corresponding rheology of the chiral material is
described by a dissipationless odd viscosity and odd elasticity~\cite{Banerjee2017,Scheibner2020}.
%
%
The study of these systems gives insight into non-equilibrium
pattern formation~\cite{Bililign2021,Kokot2017}, may be relevant for the ecology of certain organisms~\cite{Drescher2009,Petroff2018}, and provides inspiration for new types of engineering~\cite{Reichhardt2021,Brandenbourger2021}.

Collections of chiral grains moving on a vibrated substrate provide an avenue by which to reach a deeper understanding of how particle rotation at an individual level can manifest itself collectively~\cite{Tsai2005,Scholz2021}. While collections of granular rods can self-assemble to form chiral structures which spin collectively~\cite{Blair2003}, particles with tilted legs and bumpy sides, which promote frictional particle-particle interactions, have been demonstrated to spin, self-organize, and give rise to further collective motion~\cite{Altshuler2013,Koumakis2016,Scholz2018}. Particle interactions occur only during contact and present the opportunity to investigate density effects over a wide range of area fractions, in contrast to systems in which secondary flows in the interstitial medium can give rise to attraction and other system-specific effects~\cite{Drescher2009,Lopez2022}. 

\begin{figure*}
  \centering
  \begin{center}
\end{center}
  \includegraphics[width=\linewidth]{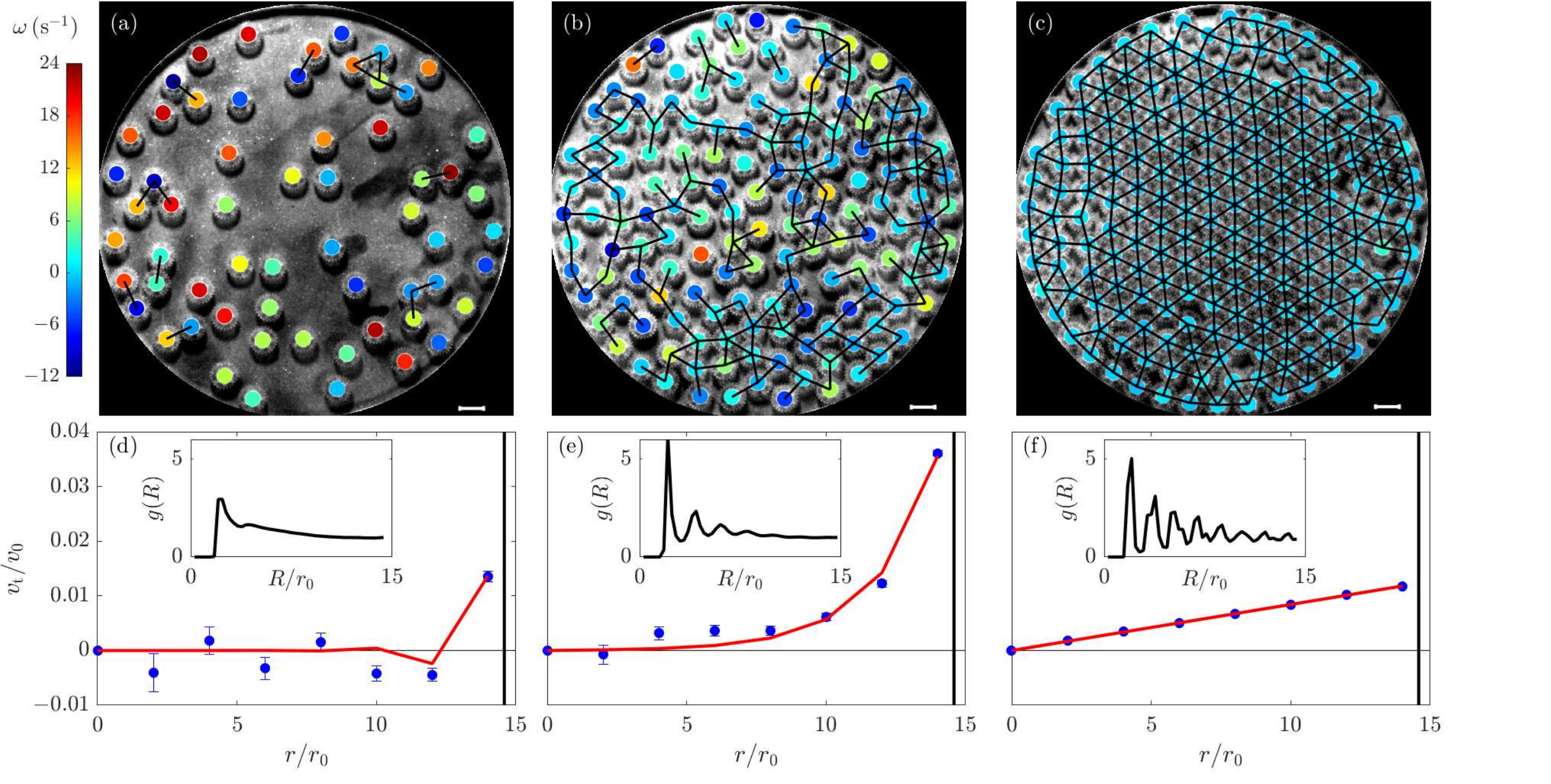}
  \caption{The dynamics of a chiral material is examined across the
    phases of gas (left column), fluid (center), and crystal
    (right). Panel (a) shows the distribution and angular velocities
    (colored spots), and contact network (solid black lines) of 60
    particles in a dilute gas ($\phi=0.23$). The scale
    bar is 1 cm.  Panels (b) and (c) show the
    corresponding particle locations, angular velocities, and contact
    networks in a fluid of 148 particles ($0.57<\phi_{\rm hex}$) and a crystal of 191 particles
    ($\phi=0.74>\phi_{\rm hex}$), respectively. Note that the mean and variance of the angular
    velocities decrease with $\phi$. The bottom row shows
    the radial velocity profile for the (d) gas, (e) fluid, and (f)
    crystal. Red lines are fits to the coarse-grained model
    (Eq.~[1--4]). The insets show the pair correlation function
    $g(R)$. The location first peak, which is insensitive to $\phi$, is found at a distance $1.08\pm0.05\,$cm. This value is between inner ($1\,$cm) and outer ($1.2\,$cm) particle diameters.}
  \label{fig:1}
\end{figure*}

Here we examine the increasing effect of particle-particle density and spin correlations on the global flow in a monolayer of chiral active matter as their area fraction
$\phi$ increases from that of a gas to a crystal. 
%
We perform experiments 
with 3D-printed particles composed of a solid gear cap 
with outer radius
$r_0=0.6$\,cm 
supported by seven elastic legs, with mass $m=0.28$\,g and moment of inertia $I=0.04\,$g$\,$cm$^2$. Because these legs are slanted, striking the particle from below causes it to spin as it accelerates upwards~\cite{Volfson2004}.  Particles are confined within a quasi-two-dimensional cylindrical
chamber of radius $R_{\rm c}=8.75\,$cm that oscillates vertically with
frequency $f=60\,$Hz and amplitude $A=0.17\,$mm. Further details on the particles and the experimental apparatus can be found in the Supplementary Document~\cite{supdoc}. 
Images are acquired at $50\,$ms intervals over ten minute time intervals. We vary the number of particles $N_{\rm p}$ from 20 to 191, corresponding to the maximum that can be achieved in the system in practice. We track the particle center and an off-center dot to measure the instantaneous position, velocity, and angular velocity of each particle in the chamber~\cite{supdoc}. 
%
%
%
%
 Isolated particles are found to
spin counterclockwise on average when viewed from above with a mean angular
velocity of $\omega_0=7.61\pm 0.01\,$s$^{-1}$. Particles diffuse with translational
diffusion coefficient $D=0.134\pm0.07\,$cm$^2$/s and rotational diffusion
coefficient $D_{\rm r}=3.4\pm0.4\,$rad$^2$/s.


Figure~\ref{fig:1}(a-c) and Supplemental Video SV1~\cite{supdoc} shows representative snapshots of the system corresponding to gas, fluid, and crystal phases with increasing $\phi$, 
where we have further superimposed the spin angular velocity $\omega$ on the tracked position of each particle. We observe that the system becomes increasingly ordered not only in the way the particles are arranged, but also in terms of their spin, with the highest $\phi$ showing hexagonal crystalline order and little variation in $\omega$. We construct the contact network from instantaneous positions of the
particles,  taking two particles to be in contact if their center to center distance is less than $0.2r_0$, and is also shown in Fig.~\ref{fig:1}(a-c).
Spanning contact networks appear around $\phi \approx 0.5$~\cite{supdoc}.
As $\phi$ is further increased, the contact
network becomes ordered and the disordered fluid becomes a crystal.

%
%
To characterize the emergence of flow and its nature, we calculate the tangential velocity of each particle as
it moves about the center of the chamber.
Averaging these instantaneous measurements over ten-minute
trials, we find the steady-state velocity profile $v_t$ as a function of distance $r$ from the chamber center.
The corresponding profiles are shown in Fig.~\ref{fig:1}{(d--f)}, nondimensionalized by the characteristic speed $v_0=r_0\omega_0=4.6\,$cm/s of an isolated particle's edge. In all cases, particle speed is maximum within a particle diameter of the chamber wall, where they move with flux $\ve= 2 r_{0} v_t$. 
The overall collective rotation in the chamber is in the same direction as the individual particle spin, which implies that the flow is driven by particle-particle collisions rather than particle-boundary interactions~\cite{Workamp2018}. The measured $v_t$ of each data set obtained between $\phi = 0.078$ and 0.746 can be found in the Supplementary Figure S5~\cite{supdoc}. 
Our results provide the first experimental investigation of flows driven by chirality as the density varies widely, corresponding to a chiral gas, fluid, and crystal states.

\begin{figure}[t]
  \centering
  \includegraphics[width=\linewidth]{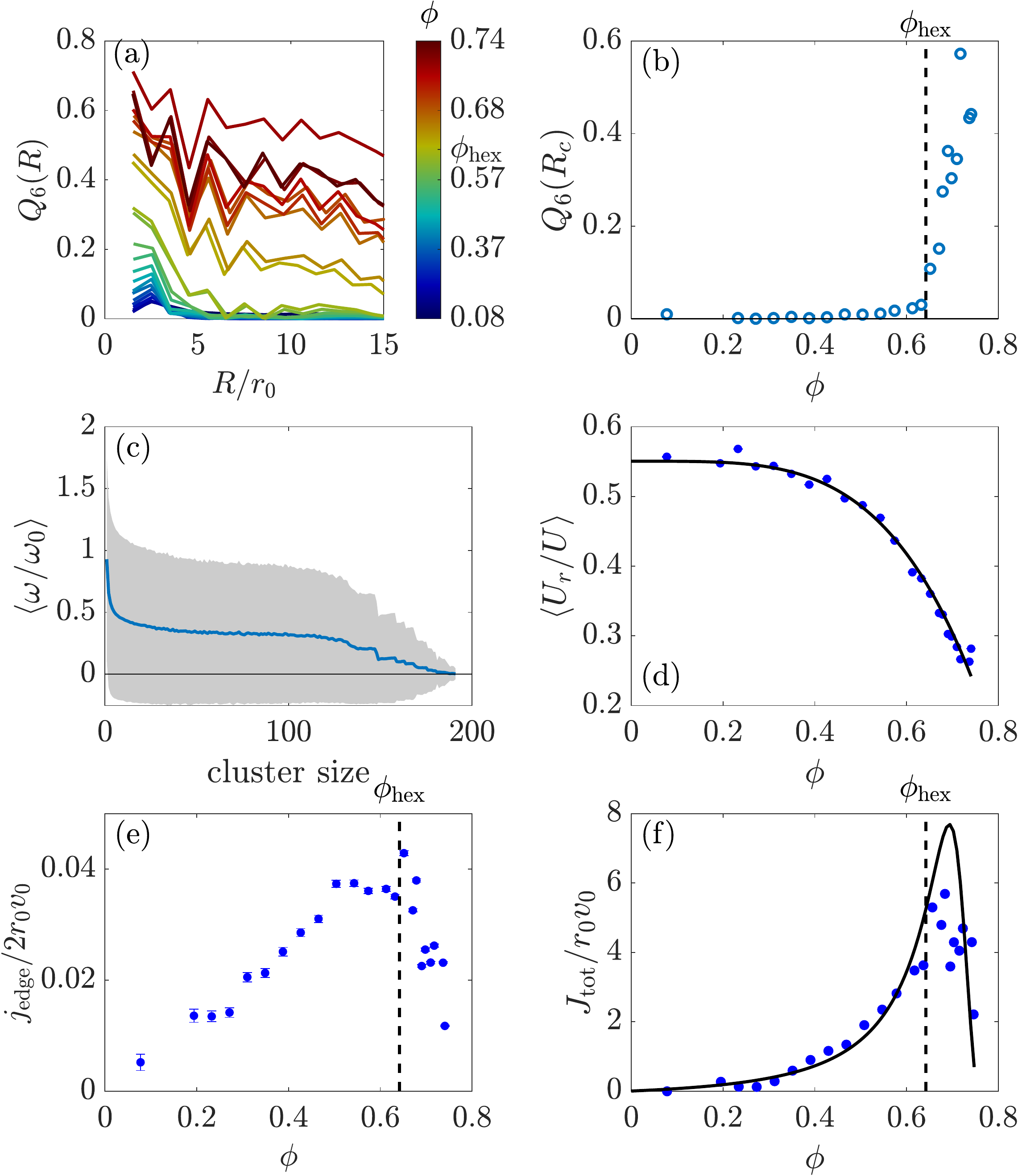}
  \caption{Cluster geometry and particle motion vary systematically with area fraction. (a) The magnitude and correlation length of the orientation of particle contacts grows with area fraction. (b) The formation of a single rotating crystal is identified from the point at which $Q_6(R_{\rm c})$ increases discontinuously. (c) Average angular velocity particles (blue line) and the typical fluctuations (shaded region) decrease as the size of the contact network grows. (d) As area fraction increases, the rotational motion is slowed more quickly than translational components. The black line, shown as a guide to the eye, is $\langle U_{\rm r}/U\rangle=0.55 - 1.03 \phi^4$. (e) The edge current $\ve$ is a non-monotonic function of $\phi$ which changes abruptly at $\phi_{\rm hex}.$ (f) The integrated particle flux $J_{\rm tot}$ is maximized at $\phi=0.69$. The black line shows the predictions of Eqs.~(1--4). }
  \label{fig:2}
\end{figure}

To quantify these transitions, we plot in Fig.~\ref{fig:2}(a) the six-fold orientational correlation
function $Q_6(R)=\langle q_6(\mathbf{R})q_6^*(0)\rangle$, where
$q_6(\mathbf{r}_k)=N_k^{-1}\Sigma_{j=1}^{N_k}e^{i 6 \theta_{k,j}}$, $\mathbf{R}$ is the position vector (of magnitude $R$) between the centers of two particles, $N_k$ is the number of particles that particle $k$ contacts, $\mathbf{r}_k$ is the position vector of the particle $k$ from the center, and $\theta_{k,j}$ is
the angle between particles $k$ and $j$, which are in contact with one
another.
As shown in Fig.~\ref{fig:2}(b), particle orientation becomes correlated at a critical area fraction of $\phi_{\rm hex}=0.64$ over the scale of the chamber $R_{\rm c}$ and the hexagonal packing of particles becomes apparent as in Fig.~\ref{fig:1}(c).
The pair-correlation function $g(R)$, which describes the density variation as a function of distance $R$ from a particle, is shown in the Insets of Fig~\ref{fig:1}(d), Fig~\ref{fig:1}(e) and Fig~\ref{fig:1}(f). They are observed to be consistent with those of a gas, fluid and crystalline solid, respectively, with the appearance of peaks growing at $R = 2r_0, 2\sqrt{3}r_0$, and $4r_0$ corresponding to a hexagonal lattice. The measured pair correlation function is shown in the Supplementary Figure S6~\cite{supdoc} for each experiment. The small secondary peak is absent at the lowest area fraction examined. 

Because colliding particles tend to rotate about one another, collisions
transfer angular momentum from the individual rotation of particles to
the global rotation about the chamber.
As the contact network grows and spatial correlations increase with $\phi$, the average spin angular velocity $\langle \omega \rangle$ and its root mean fluctuations decreases as shown in Fig.~\ref{fig:2}(c). 
We measure the particles' instantaneous two-dimensional translational
kinetic energy $U_{\rm t}=\frac{1}{2}m v_{\rm 2D}^2$, where $v_{\rm 2D}$ is the instantaneous translational speed of a particle, the rotational kinetic
energy $U_{\rm r}=\frac{1}{2}I\omega^2$,  and the total measured energy $U=U_{\rm r} +U_{\rm t}$.
%
%
As shown in Fig.~\ref{fig:2}(d), the partitioning of energy between
translational and rotational motion $\langle
U_{\rm r}/U\rangle=0.55$ in the gas phase,  and decreases quartically with $\phi$ in the fluid phase. As the packing becomes crystalline---close to $\phi_{\rm hex}$---and $\omega$ of the particles locks in phase with the solid body rotation, $\langle U_{\rm r}/U\rangle$ becomes similar to 1/3, the value predicted by the equipartition theorem.
%

%
%
In the dilute limit of a chiral gas, particle collisions are dominated
by two-body interactions.
Because the co-rotation of isotropically colliding particles generates
no net flow, the velocity field vanishes in the interior of the
chamber~\cite{Liu2020}.
The presence of the chamber wall breaks this symmetry. Because a
particle near the wall can only be struck from the chamber interior,
the outer ring of particles slip over the chamber walls as they are
pushed from the interior. Since this mechanism does not reference a particular $\phi$, one may expect an edge current even at vanishing densities.
Edge currents have been shown to develop in a numerical study of a dilute ($\phi\approx0.12$)
gas of driven rotors interacting with Yukawa potential~\cite{Dasbiswas2018}.
Figure~\ref{fig:2}(e) shows that edge currents form at area fractions as low as $0.078$ even in systems which interact sterically. We find that $j_{\rm edge}$ increases approximately linearly with $\phi$, and $j_{\rm edge}$ apparently can extend to vanishing densities provided particle-particle collisions are present.
The corresponding flux  $J_{\rm tot}= 2 \pi \int \phi v_t r \mathrm{d}r$ is shown in Fig.~\ref{fig:2}(f).
Our experiments show that a thin edge current ($2r_0$ wide)  is
maintained by short-range particle-particle interactions in a dilute gas.
Confinement induced packing structure has been shown numerically to give rise to oscillatory flows at intermediate $\phi$, but were not clearly realized in their corresponding experiments~\cite{Liu2020}. Interestingly, we observe a clear signature of oscillatory flow for  $\phi < 0.352$ (see Supplementary Figure S5~\cite{supdoc}) with a weak counterclockwise flow for $ 11 < r/r_0 < 13$ as a reaction to the clockwise edge current.     

As $\phi$ rises, the edge current and associated flux, initially grows and
extends through the system (Fig~\ref{fig:2}[e--f]).
We observe that a disordered contact network
(Fig~\ref{fig:1}(b)) maintains flow in the bulk, and thus conclude that bulk flow does not require the percolation of solid-like regions.
Rather, around an area fraction of $0.5$~\cite{supdoc}, the contact network spans the
chamber and the outermost particles cannot move independently of those
in the interior. However, the loose contact network permits the
relative motion of particles. As the outermost shell of particles is
pushed around the exterior of the chamber, it drags the loose network. In
this regime, velocity gradients begin to extend through the entire
material (Fig~\ref{fig:1}(e)).

Finally, in the dense regime, $\phi>\phi_{\rm hex}=0.64$, steric
interactions arrest the relative motion of particles, velocity
gradients are suppressed, and particles cease to rotate independently of the lattice except near defects. The amplitude of the edge current decays quickly with particle concentration (Fig.~\ref{fig:2}(e)) and the
crystal moves as a solid body (Fig.~\ref{fig:1}(f)). Interestingly, solid-body rotation is maintained even as system scale dislocations form (see Supplemental Video SV2~\cite{supdoc}). In the crystalline limit, particles rotate in phase with the solid body rotation except near topological defects (see Supplemental Video SV3~\cite{supdoc}).

\begin{figure}[t]
  \centering
  \includegraphics[width=\linewidth]{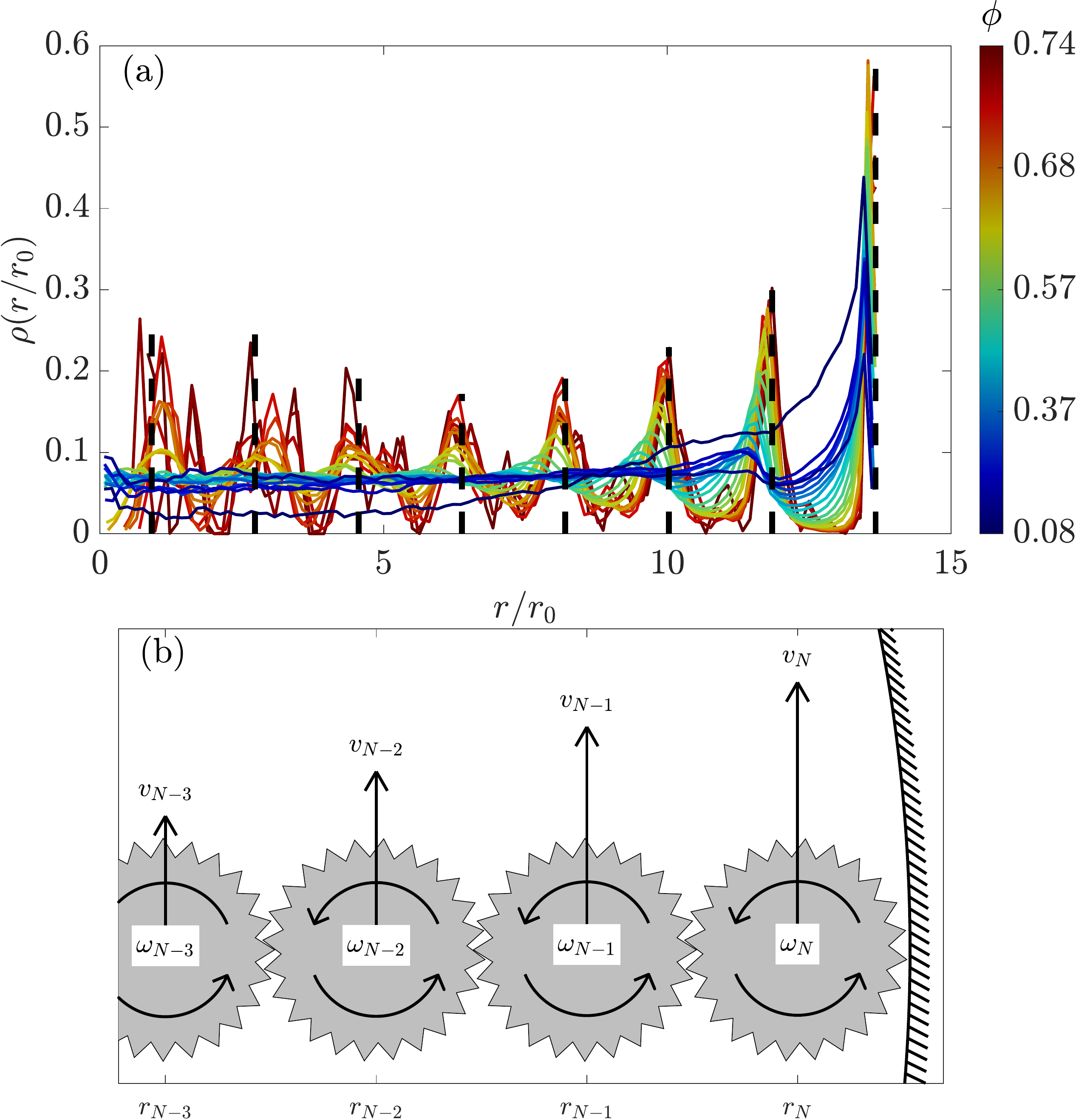}
  \caption{Particles rotate about the center of the chamber in
    concentric lanes. Lanes form from the outer boundary and
    grow inwards with increasing $\phi$. (a) The
    probability density function $\rho$ for particle density for all
    experiments analyzed is shown. The dashed lines show the expected
    locations of lanes. 
    (b) A schematic of particles interacting between lanes. Each particle moves in a circular path at velocity $v_i$ and rotates about its axis at angular velocity $\omega_i$.} 
  \label{fig:3}
\end{figure}

Next, we analyze the velocity profile quantitatively to understand
how the macroscopic flow field is reflected in the kinetics of particle
interactions.
At particle
concentrations above $\phi>\phi_{\rm hex}$, particles rotate about the
chamber in eight concentric lanes (Fig.~\ref{fig:3}[a]).
%
%
The outermost lane---at a
distance of one particle radius from the outer wall---is apparent even
at the lowest concentration examined.
Multiple lanes become apparent around the same $\phi$ at which contact networks begin to span the system (Fig.~\ref{fig:3}[a]).
We coarse-grain this system in a manner in which each
lane---having a width of one particle diameter and the outermost lane
is centered one particle radius from the outer wall---is densely
occupied with the average particle concentration as illustrated Fig.~\ref{fig:3}(b).
%


Consider the torque balance on particles in the $i^{\rm th}$ lane from
the center with velocity $v_i$ and spin angular velocity $\omega_i$, where $i<N$ and $N\approx R_{\rm c}/(2r_0)$ is the number of lanes that fit in the experiment.
The rotation of these particles is slowed at rate $\alpha_p$ if the
speed of its edge is faster than those of its neighbors.
The corresponding nondimensionalized torque balance on the particles,
as derived in Supplemental Document~\cite{supdoc}, is
\begin{equation}
  \omega_i+\dfrac{\alpha_p}{\alpha_b}\left( v_{i-1} -v_{i+1} +\omega_{i-1} +4\omega_i +\omega_{i+1}\right)=1,
  \label{eq:omega_i}
\end{equation}
where $\alpha_b$ is the rate that particle rotation is slowed by the
bottom of the chamber.
In the outermost lane, particle rotation is only slowed by collisions
from the interior, within its lane, and the bottom of the chamber.
The boundary condition on particle rotation at the wall is
\begin{equation}
  \omega_N+\dfrac{\alpha_p}{\alpha_b}\left( v_{N-1} -v_{N} +\omega_{N-1} +3\omega_N\right)=1.
  \label{eq:omega_N}
\end{equation}
We assume that $\omega$ is smooth and continuous near
$r=0$.
%

Similarly, $v_t$ is slowed at rate $\beta_p$ if a particle's edges move more quickly than the edges it contacts.
The corresponding torque balance on lane $i<N$ requires
\begin{equation}
  v_i=\dfrac{\beta_p}{\beta_b}\left(v_{i-1}-2v_i +v_{i+1}
  -\omega_{i+1}+\omega_{i-1}\right),
    \label{eq:v_i}
\end{equation}
where $\beta_b$ is the rate that the translational velocity is slowed
by the bottom of the chamber. Again, the outermost particles are only
affected by particles in the lane centered at $r_{N-1}$. The
corresponding boundary condition is
\begin{equation}
  v_N=\dfrac{\beta_p}{\beta_b}\left(v_{N-1}-v_{N}+\omega_{N-1}+\omega_{N}\right).
    \label{eq:v_N}
\end{equation}
The boundary condition at the center of the chamber requires the
angular velocity $\Omega=v_t/r$ to be smooth and continuous. 

Equations (\ref{eq:omega_i}--\ref{eq:v_N}) uniquely determine the
tangential velocity and the angular velocity of particles in each
lane.
We fit the dimensionless relaxation rates $\alpha=\alpha_p/\alpha_b$ and $\beta=\beta_p/\beta_b$ to match the measured velocity profile. Three representative fits are shown in
Fig~\ref{fig:1}(d--f), and all the trials are shown in the Supplemental Figure S5~\cite{supdoc}.
Remarkably, this model reproduces the velocity profile even in the
dilute regime and correctly predicts the slight retrograde motion in
the $N-1$ lane (Fig~\ref{fig:1}(d) and Fig.~S5).
%

\begin{figure}[t]
  \centering
  \includegraphics[width=\linewidth]{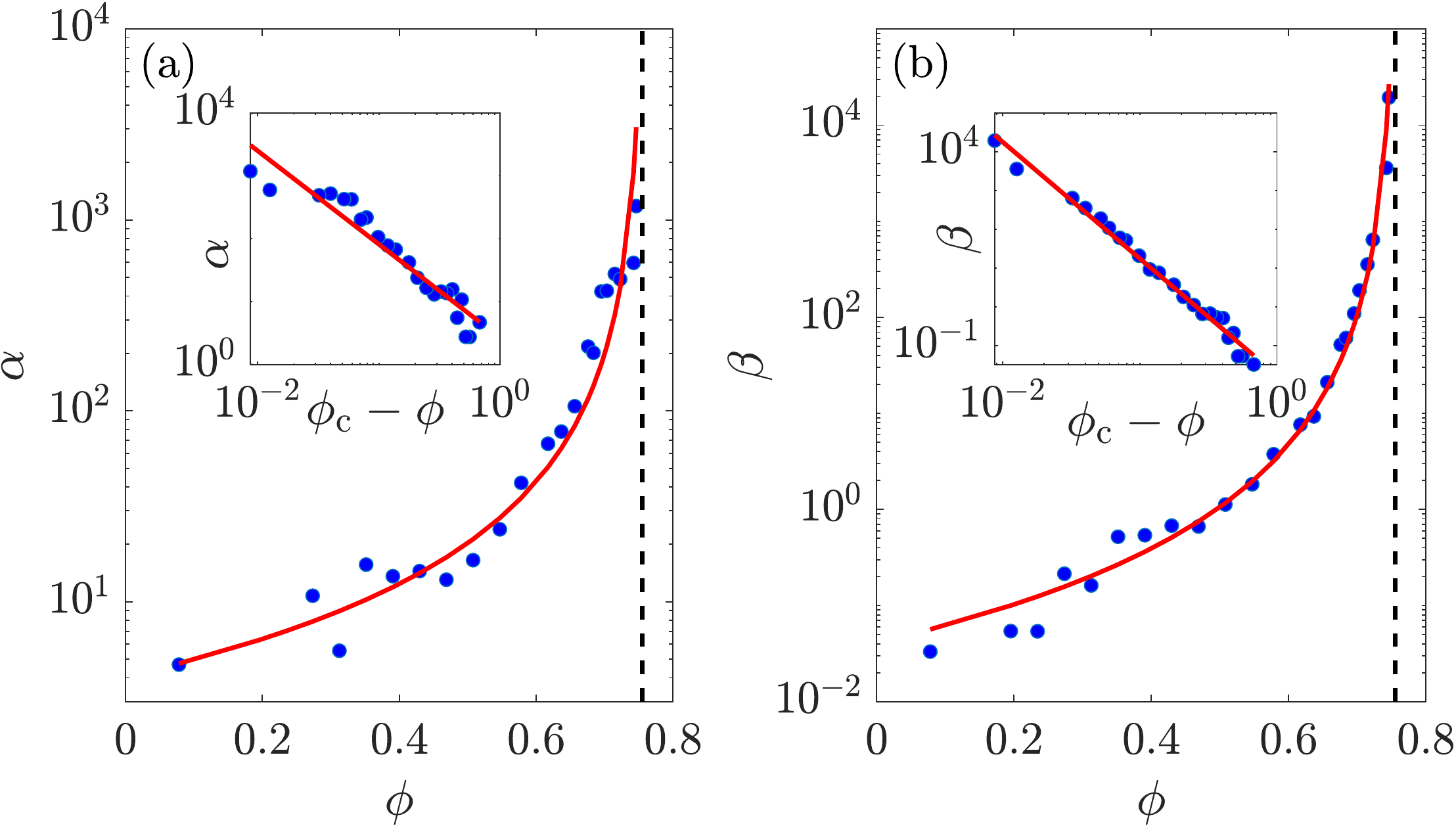}
  \caption{The rates at which particle (a) angular velocity and (b)
    tanslational velocity relax to the speeds of the neighboring edges
    both diverge at the maximum area density. The red lines are
    fits to power laws. The insets show the same data on
    logarithmically scaled axes.}
  \label{fig:4}
\end{figure}

Intuitively, the relaxation rates $\alpha$ and $\beta$ should increase
with particle density as increasing the number of collisions similarly
increases the rate particles are slowed by their neighbors.
As shown in Fig.~\ref{fig:4}, $\alpha$ and $\beta$ increase faster
than exponentially with area fraction.
%
%
These trends are well fit by power law divergences
$\alpha(\phi)=\alpha_0 (\phi_{\rm c}-\phi)^{-\gamma/2}$ and
$\beta(\phi)=\beta_0 (\phi_{\rm c}-\phi)^{-\gamma}$, where
$\alpha_0=1.53$, $\beta_0=0.17$, $\gamma=3$, and $\phi_{\rm
 c}=0.76$.

Figure~\ref{fig:2}(f) shows that the particle flux  predicted by the power law divergences of $\alpha$
and $\beta$ approximates the measured flux reasonably well.
The predicted flux is not monotonic~\cite{yang2020robust} and vanishes at a value of $\phi_{\rm c}=0.76$. 
%
The predicted flux is maximized at $\phi \approx 0.69$ and is similar to the value $\phi_{\rm s}=0.711$ at which a large two-dimensional lattice of hard spheres transitions from diffusive behavior to caging, as discussed by Reis, et al.~\cite{Reis2007}.
The slightly lower value could result from the difference in particle shapes and finite size effects. 
This similarity suggests that flux is maximized when the rate of particles collisions is maximized before the relative motion of particles is arrested.

In conclusion, we have analyzed the evolution of edge current and bulk flow across three phases of active chiral matter. Edge currents are observed even at vanishing densities due to occasional particle collisions and shielding of particles at the boundaries.
Upon the onset of system-scale orientational order, the edge current is quickly arrested and particles rotate as a solid body.
A coarse-grained model, which respects the emergent crystalline order and the finite particle size, accurately fits the measured velocity profile
across these phases.

\begin{acknowledgments}
We thank Trinh Huynh and Animesh Biswas for help with building experimental components, and J\"orn Dunkel for bringing Ref.~\cite{Dasbiswas2018} to our attention. This work was partially supported by NSF Grant no. DMR-2005090 and NSF Grant no. PHY-2042150.
\end{acknowledgments}

\bibliographystyle{unsrt}

\end{document}